\documentclass[aip,apl,reprint,preprintnumbers]{revtex4-1}

\usepackage{CJK}

\usepackage{graphicx}
\usepackage{dcolumn}
\usepackage{bm}
\usepackage{color}
\usepackage{sidecap}

\begin{document}

\title{Multi-gigahertz operation of photon counting InGaAs avalanche photodiodes}
\author{Z. L. Yuan}
\email{zhiliang.yuan@crl.toshiba.co.uk}
\author {A. W. Sharpe}
\author {J. F. Dynes}
\author{A. R. Dixon}
\altaffiliation[Also at ] {Cavendish Laboratory, University of Cambridge, J. J. Thomson Avenue, Cambridge CB3 0HE, UK.}

\author {A. J. Shields}

\affiliation{Toshiba Research Europe Ltd, Cambridge Research
Laboratory, 208 Cambridge Science Park, Milton Road, Cambridge, CB4~0GZ, UK }

\date{\today}

\begin{abstract}
We report a 2~GHz operation of InGaAs avalanche photodiodes for efficient single photon detection at telecom wavelengths. Employing a self-differencing circuit that incorporates tuneability in both frequency and arm balancing, extremely weak avalanches can now be sensed so as to suppress afterpulsing. The afterpulse probability is characterized as 4.84\% and 1.42\% for a photon detection efficiency of 23.5\% and 11.8\%, respectively. The device will further increase the secure bit rate for fiber wavelength quantum key distribution.
\end{abstract}


\maketitle

Semiconductor InGaAs avalanche photodiodes (APDs) offer a practical solution for single photon detection in the near infrared, because they are rugged and compact, cryogen-free and of low power consumption.
Due to their practicality, InGaAs APDs are widely used in a number of important applications, such as quantum key distribution,\cite{dixon08} as well as eye-safe laser ranging,\cite{maruyama02} and semiconductor device characterization.\cite{ward05}
Despite their afterpulse problem, the photon counting performance has been steadily improved over the past decade. Under gated mode, the gating frequency has been increased from initially 100~kHz to presently over 1~GHz due mainly to advances in detection circuitry.\cite{bethune00,tomita02,yoshizawa04,namekata06,yuan07} A count rate of 497~MHz has recently been demonstrated.\cite{dixon09}

Detecting weak avalanches is the key to high speed single photon detection. In gated mode, weak avalanches are usually not immediately accessible because of capacitive response by the APD to the gate signal. They can be recovered through removal of the capacitive response, for which both sine-wave gating\cite{namekata06} and self-differencing (SD)\cite{yuan07} have been demonstrated as highly effective. Sine-wave gating relies on effectively filtering the capacitive response that is purely sinusoidal,
while SD removes the capacitive signal by subtracting an identical copy that is delayed by an integer number of clock cycles. Contrary to the sine wave gating, any form of gating, including sine waves\cite{zhang09b} and low-duty cycle pulses,\cite{xu09} is acceptable by an SD as long as it is periodical. Through SD or sine-wave gating, gigahertz operation \cite{yuan07,namekata09} has been achieved with their respective performances summarized in Table~\ref{tab:comparison}. InGaAs APDs have been applied to Mbits/s secure key rate quantum key distribution,\cite{dixon08} generation of high quality random numbers,\cite{dynes08} and long distance entanglement distribution.\cite{dynes09}

\begin{table}[t]
\caption{Comparison of the present work with previously reported performances for gigahertz-clocked InGaAs APDs. $f$: frequency, $\eta$: detection efficiency, $P_a$: afterpulse probability, and $P_d$: dark count probability.}
\begin{ruledtabular}
\begin{tabular}{lcccc}
  &$f$(GHz) &$\eta(\%)$ &$P_a$(\%) & $P_d$(gate$^{-1}$)\\
\hline SD-APD, 2007\footnotemark[1] & 1.25 
& $10.9$& 6.16 & $2.34\times10^{-6}$\\
\hline Sine-wave, 2009\footnotemark[2] & 1.5 
&  10.8 & 2.8 & $6.3\times10^{-7}$ \\
\hline This work  & 2.0 
& 11.8 & 1.43 & $3.79\times10^{-6}$\\
 & 2.0
 & 23.5 & 4.84 & $1.32\times10^{-5}$\\
\end{tabular}
\end{ruledtabular}
\footnotetext[1] {Reference~\onlinecite{yuan07}.}
\footnotetext[2] {Reference~\onlinecite{namekata09}.}
\label{tab:comparison}
\end{table}

\begin{figure}[b]
\centering\includegraphics[width=1\columnwidth]{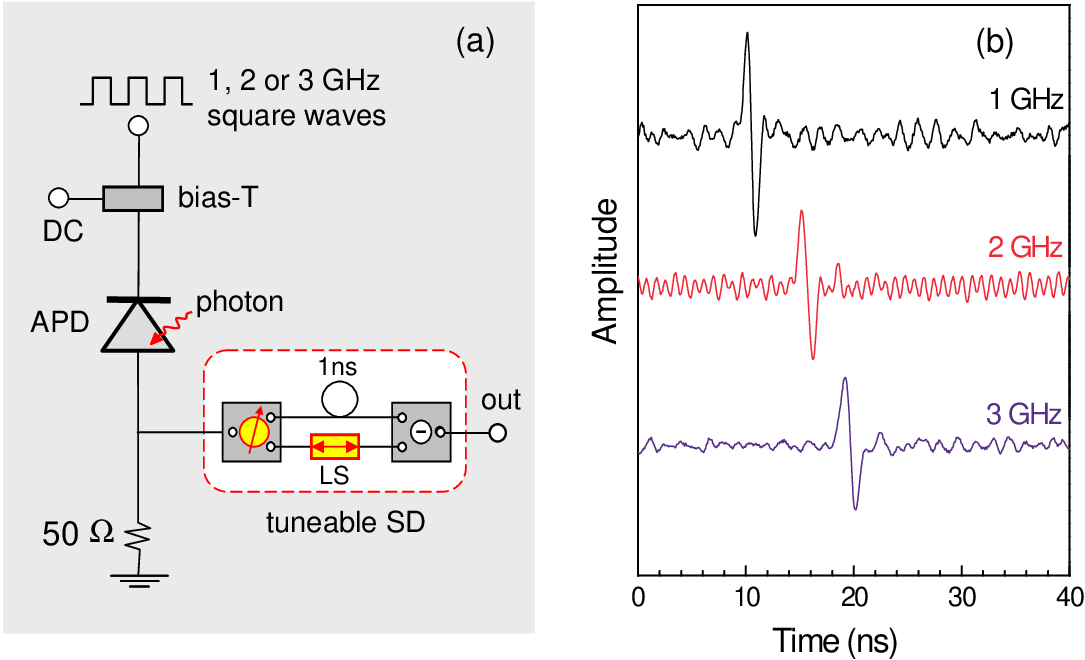}
\caption{(a) Schematic for a tunable SD circuit with 1-ns delay; LS: line-stretcher. (2) Avalanche waveforms for dark events recorded by an oscilloscope after the 1-ns SD circuits under different gating
frequencies. The waveforms are plotted in the same scale but shifted vertically for clarity.}
\label{fig:sd}
\end{figure}

As pointed out in Ref. \onlinecite{yuan07}, continuous frequency tuning was not possible in the original SD circuit, preventing an ideal performance due to the difficulty in perfect frequency match to a given photon source.  This inflexibility can be totally relieved by
adding adjustability into the circuit, as shown in Fig.~\ref{fig:sd}(a). In comparison to the original SD,\cite{yuan07} the tuneable circuit
replaces the 50/50 splitter with an adjustable splitter and also a co-axial line stretcher.  Adjustability in the splitting ratio is realized by adding a potentiometer in a triangular resistive splitter. The line stretcher has a tuneable delay range of 45~ps, thus allowing the SD to match a frequency within the range from 0.987 to 1.033~GHz with an arbitrarily high precision. Thanks to the adjustability in splitting ratio and frequency, the SD circuit allows greater suppression of the capacitive response. The suppression is measured to be 62~dB for the 1~GHz frequency components. The overall cancelation\cite{sd} for the capacitive response is approximately 38~dB, which is 17~dB better than the non-tuneable SD circuit.\cite{yuan07}

In what follows, we report a significantly improved performance of InGaAs APDs due to the tuneable SD. A gating frequency of 2~GHz operation is demonstrated for single photon counting with a high detection efficiency of 23.5\% and a low afterpulse probability of 4.85\%. As compared with our previous results (see Table~\ref{tab:comparison}), the present detector operates
at almost twice the frequency and twice the efficiency.

The InGaAs APD under test is cooled electrically to $-30^\circ$C. A square wave in combination with a DC voltage source is used as the gating signal to bias the device, as illustrated in Fig.~\ref{fig:sd}(a). The APD signal, after passing through
the SD, is amplified before analysis by an oscilloscope or a time-correlated photon counter. The electrical current through the DC path is
monitored.

The superiority of the tuneable SD is visible in the traces of avalanches that are due to dark events.  The use of dark events helps to avoid complications due to the avalanche amplitude dependence on photon number,\cite{kardynal08} which may give rise to a false confidence in
the circuit performance. Figure~\ref{fig:sd}(b) shows three dark events obtained with a 1-ns delay SD when the APD is gated at a frequency of 1, 2 and 3 GHz, respectively. For all frequencies, the SD shows an excellent level of cancelation. The uncanceled background is much weaker than the avalanche signal, which can be straightforwardly discriminated.

\begin{figure}[t]
\centering\includegraphics[width=.75\columnwidth]{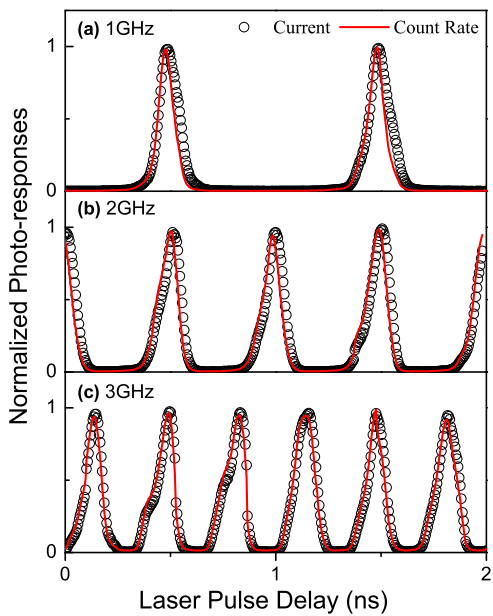}
\caption{ Photon current and count rate as a function of the laser pulse delay measured by an InGaAs APD under a square wave gating of (a) 1~GHz, (b) 2~GHz, and (c) 3~GHz.}
\label{fig:delay}
\end{figure}

\begin{figure}[b]
\centering\includegraphics[width=.8\columnwidth]{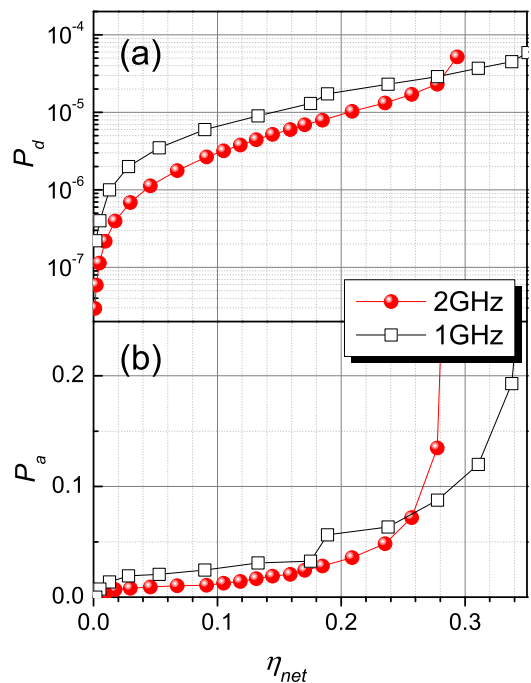}
\caption{ (a) Detector dark count probability ($P_d$), and (b) afterpulse probability ($P_a$) as a function of the
net detection efficiency ($\eta_{net}$).}
\label{fig:performance}
\end{figure}

For high speed gating to be useful, it is important for an APD \emph{not} to react strongly to photons arriving between gates, but only be sensitive to those arriving within an electrical gate. To verify this, we illuminate the
APD with a 1~GHz pulsed laser whose intensity is attenuated to 3~pW, corresponding to 0.023 photons/pulse. The APD gating is synchronized with
the laser at its harmonic frequencies of 1, 2, and 3~GHz. Photocurrent,  as well as photon count response, is recorded as the laser pulse delay is varied relative to the gating.  As shown in Fig.~\ref{fig:delay}, both responses show very similar behavior: the measured values vary with the laser pulse delay, forming a series of peaks whose frequency corresponds well with the respective APD gating period.  These peaks are attributed to the delays with which the incident photons coincide with the APD gates.
Under such conditions, each absorbed photon have a high probability of inducing a macroscopic avalanche. When the laser pulse is tuned between the gates, the photo responses, including photocurrents, fall close to their dark values, suggesting that the off-gate photons have a negligible avalanche gain or detection probability. The full width half maximum (FWHM) for each response peak is approximately 100~ps, showing little dependence on the gating frequency.

To characterize the photon counting performance, the APD is illuminated with 50-ps optical pulses from a 1550~nm semiconductor laser diode
synchronized at 1/64 of the APD gating frequency. The laser power is attenuated to 0.128~pW, corresponding a photon flux of $\mu=1.0\times10^6$~s$^{-1}$, before launching into the fiber pigtail of the APD. At 1 and  2~GHz gating, each illuminating pulse contains
0.064 and 0.032 photons, respectively, thus ensuring operation in the single photon counting regime. By fixing the gating amplitude at 7.1~V, we characterize the raw count rate ($C$), dark count ($P_{d}$), and afterpulse ($P_{a}$) probabilities as function of the APD DC voltage
bias, using the method described previously.\cite{yuan07} We point out that the dark count probability is measured per gate, while
the afterpulse probability is measured per photon count. Excluding the dark and afterpulse counts, the net detection efficiency ($\eta_{net}$) is obtained using
\begin{equation}
\eta_{net}=\frac{(C-P_{d}f)/\mu}{1+P_{a}},
\end{equation}
\noindent where $f$ is the clock frequency.

The characterization results, plotted in Fig.~\ref{fig:performance},  shows a better performance for a gating frequency of 2~GHz than 1~GHz.  For similar detection efficiencies, the APD shows much lower values for both $P_d$ and $P_a$ for 2~GHz operation. The detection efficiency is considerably better than previous results.\cite{yuan07}
For $P_a<10\%$, the net efficiencies are 27.8\% ($P_a=8.8\%$) and 25.7\% ($P_a=7.2\%$) for 1 and 2~GHz respectively. The corresponding dark count probabilities are $2.90\times10^{-5}$ and $1.71\times10^{-5}$.
The afterpulse probability is also significantly smaller. At $\eta_{net}=11.8\%$, a typical efficiency for InGaAs APDs, the afterpulse probability is measured to be 1.43\% for 2~GHz. This value is four times lower than our previously reported value\cite{yuan07} for 1~GHz, and is also 50\% lower than the recent value\cite{namekata09} reported for a 1.5~GHz sine-wave gating APD. At an elevated efficiency of $\eta_{net}=23.5\%$, $P_a$ only increases slightly to 4.84\%. The reduced afterpulse probability is advantageous for error-sensitive applications, such as quantum key distribution.

\begin{figure}[t]
\centering\includegraphics[width=.87\columnwidth]{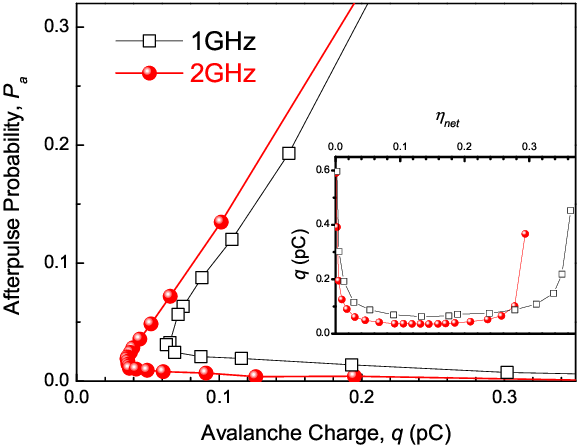}
\caption{Afterpulse probability ($P_a$) \textit{vs.} avalanche charge ($q$). The inset shows the avalanche charge as a function of the net
detection efficiency ($\eta_{net}$).}
\label{fig:charge}
\newpage
\end{figure}

The reduced afterpulse probability agrees well with the reduction in the avalanche charge $q$. The relation between $P_a$ and $q$ is plotted in Fig.~4. The value of $q$ is obtained using the ratio of photo-current to count rate,\cite{yuan09} representing the total charge per detection event that flows through the device.  Two distinctive relations are observed:  (i) $P_a$ that decreases with $q$ and (ii) a linear relation.  The former is an artifact, because these data are obtained with low efficiency (see inset of Fig.~\ref{fig:charge}), or low bias conditions, under which $q$ is highly overestimated because a considerable fraction of avalanches are not sufficiently strong for detection.  The latter is the actual relation, which explicitly reveals that the afterpulse probability increases linearly with the avalanche charge. The lowest $q$ is measured as 0.068~pC for 1~GHz. The value is less than 50\% of a previous value measured at a similar frequency.\cite{yuan09} For 2~GHz, $q$ becomes even lower with a value of 0.035~pC recorded. This significant reduction in $q$ helps the APD to achieve high efficiency with minimal afterpulse at high frequencies, and is the origin of the lower afterpulse rate at higher gating frequencies.

As summarized in Table~\ref{tab:comparison}, the present APD operates almost double the frequency and also double the efficiency, compared to our previous result.\cite{yuan07} Moreover, the higher clock frequency does not require a more stringent synchronization with the laser.  Based on the detector response curves in Fig.~\ref{fig:delay},
synchronization with the laser requires the same precision at 2~GHz as 1~GHz, which is already demonstrated in high bit rate quantum key distribution\cite{dixon08} and long distance entanglement distribution.\cite{dynes09} As a result, the improved performance of the detector is expected to allow a four-fold increase in the secure bit rate of quantum key distribution.

An outstanding issue remains to be addressed for multi-gigahertz gating.  For this APD, the timing jitter is measured as 100, 120 and 380~ps for 1, 2 and 3~GHz gating, respectively, showing an increasing trend with gating frequency. The amount of jitter is acceptable for 1 and 2~GHz, but is prohibitively high for 3~GHz because it is considerably greater than the clock period of 333~ps at this frequency. The jitter, and thus the maximum gating frequency, of the detector is limited by the bandwidths of the electronics, the diode itself, or a combination of both. In the future it is intended to employ devices, as well as electronics, of higher bandwidths to circumvent this issue.

In conclusion, we have demonstrated high efficiency single photon detection at telecom wavelengths using self-differencing InGaAs APDs operated at multi-gigahertz gating frequencies. The device shows a detection efficiency of 23.5\% when operating at 2~GHz and with an afterpulse
probability of 4.84\%. The enhanced efficiency and speed will significantly increase the bit rate of quantum key distribution.


\begin{thebibliography}{10}
\bibitem{sd}%
  \BibitemOpen
  \bibinfo {note} {The cancelation is defined here as $20\log\frac{v_2}{v_1}$,
  where $v_2$ and $v_1$ are peak-peak amplitudes for the remaining capactive
  response after the SD with one arm broken or both arms attached
  respectively}\BibitemShut{NoStop}%
\end{thebibliography}

%

\end{document}